# Combined concurrent Physical and Chemical model for accelerated weathering damages of polyurethane-based coatings

Ambesh Gupta[1], Soumyadipta Maiti[1], Parvesh Saini[1], Surabhi Srivastava[1], Shankar Kausley[1], Beena Rai[1]

[1]TCS Research, Tata Consultancy Services Limited, Plot No. 2 & 3, MIDC-SEZ, Rajiv Gandhi Infotech Park, Hinjewadi Phase III Pune, 411057, Maharashtra, India

## Abstract

Paints and coatings undergo a variety of physical and chemical changes under environmental exposures. Accurate prediction of these changes is important for the applications of coatings. This work presents a novel approach to modeling accelerated weathering of coating by combining concurrent physical and chemical processes. The model integrates key factors influencing coating degradation and employs a multi-scale framework to capture macro-scale physical changes and micro-level chemical transformations. The chemical component/model simulates photo-degradation reactions using kinetic equations, while the physical component uses Monte Carlo simulations where repeated random events develop surface erosion. The surface topography and chemistry of coating is generated statistically through the physical model. The variations in surface topography and chemistry of coating are correlated with the chemical changes from the chemical model, resulting in estimation of both physical and chemical changes in the coating during real accelerated weathering time. Results demonstrate accurate predictions of chemistry changes, surface degradation profiles, roughness, gloss loss, and relative fracture toughness, which are validated successfully with the experimentally available data. This integrated approach provides insight into coating failure mechanisms, enabling accurate service life prediction, and serve as a tool for formulating durable coatings and optimizing testing protocols.

**Keywords:** Coating degradation, Accelerated weathering, Chemical kinetics, Durability, Surface Roughness, Predictive monitoring

# Nomenclature

| Symbol | Description | Symbol | Description |
|---|---|---|---|
| $X-X$ or $XH$ | impurity | $D_{O_2}$ | diffusivity of oxygen in polyurethane film, $m^2/s$ |
| $X^*$ | impurity radical | $P_{O_2}$ | partial pressure of oxygen in air, $Pa$ |
| $RH$ | main polymer | $E_0$ | radiation intensity of light at coating surface, $W/m^2$ |
| $R^*$ | polymer radical | $\alpha$ | Spectral sensitivity of hydrogen abstraction |
| $O_2$ | oxygen | $\varphi$ | Quantum yield of polyurethane resin |
| $ROO^*$ | peroxy radical | $A$ | Absorbance of the specie/component |
| $ROH$ | alcohol | $K$ | Constant |
| $R_2CONH_2$ | ketone | $c(y)$ | Concentration of the specie as a function of depth, $mol/m^3$ |
| $R_2COOH$ | carboxylic acid | $N_p$ | Number of photons |
| $RO$ | urethane product | $h$ | Planck's constant, $J.s$ |
| $R-R$ | crosslinked polymer | $v_{UV-A}$ | Frequency of UV-A radiation, $m^{-1}$ |
| $R_1, R_2$ | remaining part of polymer | $p$ | probability of photon absorption by material in cell |
| $t$ | time, $s$ | $\lambda$ | Wavelength of incident light, $m$ |
| $y$ | depth, $m$ | $R_a$ | Average roughness, $m$ |
| $l$ | thickness of coating, $m$ | $R_t$ | Maximum height of the surface profile, $m$ |
| $\partial/\partial t$ | differential with respect to time | $R_z$ | Average maximum height of the surface profile, $m$ |
| $k_1$ | rate constant for initiation reaction of radical from impurities (R1), $m^2/W.s$ | $x$ | Distance over coating width, $m$ |
| $k_2$ | rate constant for propagation reaction to urethane linkage (R2), $m^3/mol.s$ | $w$ | Coating width, $m$ |
| $k_3$ | rate constant for formation reaction of per-oxy radical (R3), $m^3/mol.s$ | $h(x)$ | Height of coating at $x$ distance in width, $m$ |
| $k_4$ | rate constant for termination reaction of per-oxy (R4), $m^3/mol.s$ | $E$ | Young's modulus, $Pa$ |
| $k_5$ | rate constant for per-oxy reaction with polyurethane (R5), $m^3/mol.s$ | $\gamma_s$ | Surface free energy per unit area, $J/m^2$ |
| $k_6$ | rate constant for cross-linking reaction (R6), $m^3/mol.s$ | $a$ | Maximum flaw size, $m$ |
| $S_{eq}$ | solubility of oxygen in polyurethane film, $mol/m^3.Pa$ | $\sigma_G$ | Critical stress at which crack propagates, $Pa$ |

## 1. Introduction

Exterior coatings have been widely used as a functional material on automobiles, civil structures etc. to improve surface finish, provide protection from environmental exposure, provide gloss, etc. For the coating functional materials, a high service lifetime is desired. However, the coatings can degrade over time under environmental exposure due to factors such as UV radiation, moisture, temperature fluctuations, and chemical exposure. Chemical and physical modifications to the coatings significantly influence their service lifetime. It is possible for a coating to meet the chemical requirements but fail in terms of its physical requirements. Therefore, it is crucial to estimate both physical and chemical modifications due to external weathering conditions. Knowledge of gradual time dependent failure of the coatings or the changes, it undergoes during its lifetime becomes essential for appropriate applications of coatings [1-3].

Traditional testing protocols / methods to predict coating degradation often rely on empirical data and experimental observations. For example, Florida natural exposure testing is carried out by exposing coated samples to actual outdoor conditions for 5-10 years to estimate the service lifetime of the coating [4-6]. While these tests provide realistic assessment but are time-consuming and have limited control over environmental variables. Thus, it is impossible to develop and access the coatings in different environments for a variety of applications. In recent times, Coating formulators have heavily relied on accelerated weathering tests to simulate severe outdoor weathering conditions in controlled laboratory settings. Accelerated weathering tests have reduced the testing times to 3-8 months. However, these testing protocols might fail to capture the actual physics in real-world conditions [4,7-9]. To overcome the limitations associated with these destructive testing protocols, advancements in computational modeling have been leveraged recently to accurately estimate the coating degradation in much reduced time ranging from few days to few weeks.

The computational modeling techniques utilizes robust mathematical models and simulations to accurately predict the coating degradation. Additionally, these models can significantly accelerate the pace of coating development and optimization for various applications. The robust mathematical models are either based on stochastic approach or quantitative mechanistic models [10-11]. The stochastic / statistical approach simulates the photodegradation process as a combination of small and rare events. While the quantitative mechanistic models utilize mathematical descriptors derived from the degradation mechanisms to simulate the damage.

Researchers have developed various computational models to understand the mechanisms of coating degradation and predict their performance. Early work in this area includes a study by Martin *et al.* [12], who predicted the service life of poly (methyl methacrylate) films using a stochastic model. The model uses Poisson distribution in determining probability of chain scissions and probability of a performance parameter of coating, known as the dosage and damage functions, respectively. However, these functions require extensive laboratory exposure testing to calibrate such a stochastic model. In yet another study, Hinderliter *et al.* [13] took a different approach by using a statistical Monte Carlo simulation to model coating degradation, which was represented as surface erosion from repeated degradation events. The model qualitatively reproduced the behavior of coating throughout the lifetime in terms of wetting, reflectance, and fracture toughness. Although these statistical models effectively captured the physical modifications, they lacked the ability to estimate the chemical composition changes due to the absence of degradation mechanisms. Additionally, these studies relied on arbitrary timestep units in their models that were not directly correlated with real-world weathering times.

More recently, researchers have focused on investigating coating degradation based on the chemical degradation pathways using mechanistic models. Kill *et al.* [14] developed a mathematical model by dividing the degradation process in continuum of small simultaneous steps, where each step solves several equations of the degradation process. Using this approach, the thickness of surface oxidation zone was predicted depending on the exposure and coating chemistry. In another study, Makki *et al.* [15] combined a kinetic Monte Carlo simulation with the Dissipative Particle Dynamics method, enabling the model to account for both chemical and physical pathways in the photo-degradation process. However, the study ignored the contributions of oxidation chemistry, evaporation of small molecules. Similarly, Adema *et al.* [16] employed a coarse-grained simulation method based on the kinetic Monte Carlo scheme to estimate the depth-resolved changes in optical properties and chemical composition of the polyester-urethane coatings. While these mechanistic models effectively predicted variations in chemical composition, they struggled to estimate the physical properties of coating under degradation.

This work presents a novel, combined model for predicting both the physical and chemical performance of coating under accelerated weathering exposure. A mechanistic model, grounded in chemical kinetics predicts the variation in chemical composition and provides insights into the reaction mechanisms, and degradation rates. The calibrated kinetics model using experimental data, inputs parameters to a statistical model employing a Monte Carlo scheme to simulate the physical degradation. Statistical model offers a handful of insights into the variation in the physical properties of the coating. By correlating the chemical composition obtained in mechanistic model to the statistical model, we achieve a comprehensive prediction of physical and chemical properties of the coating under real-world weathering conditions. This combined model enables accurate service life predictions based on material's evolving chemical and physical properties. Ultimately, this combined model not only deepens on the predictions of coating degradation but also empowers the development of more robust and durable coating for various applications.

The mechanistic model is referred to as kinetics model, while statistical model is referred as Monte Carlo model in the further sections of this study. This study is organized into 4 sections. Section 2 describes the methodology utilized in developing the kinetics model and its optimization procedure. Also, the methodology to develop and simulate the Monte Carlo model is presented in the same section. Results and Discussion from the combined model to predict the degradation behavior of coating are presented in section 3. In Section 4, key insights from the combined model are summarized, along with the future scope of this work.

## 2. Methodology

Most of the degradation studies in the past have been carried out on acrylic-melamine, epoxy-amine coating for automobiles [4,14-16]. However, an acrylic-polyurethane coating chemistry is the most widely used in automobile industry as a protective exterior coating [17]. Thus, acrylic-polyurethane coating system is the most relevant system to be investigated for validation of mechanistic models in this work.

A. Mechanism of polymer chemical degradation

The mechanism of photodegradation for acrylic-polyurethane coating is a complex process that includes several steps and can be influenced by several factors such as specific formulation of coating, UV radiation intensity, environmental factors like temperature, humidity, etc. Based on the detailed studies of Larche [18], Morrow [19], following reduced mechanism is identified for the photodegradation in this work.

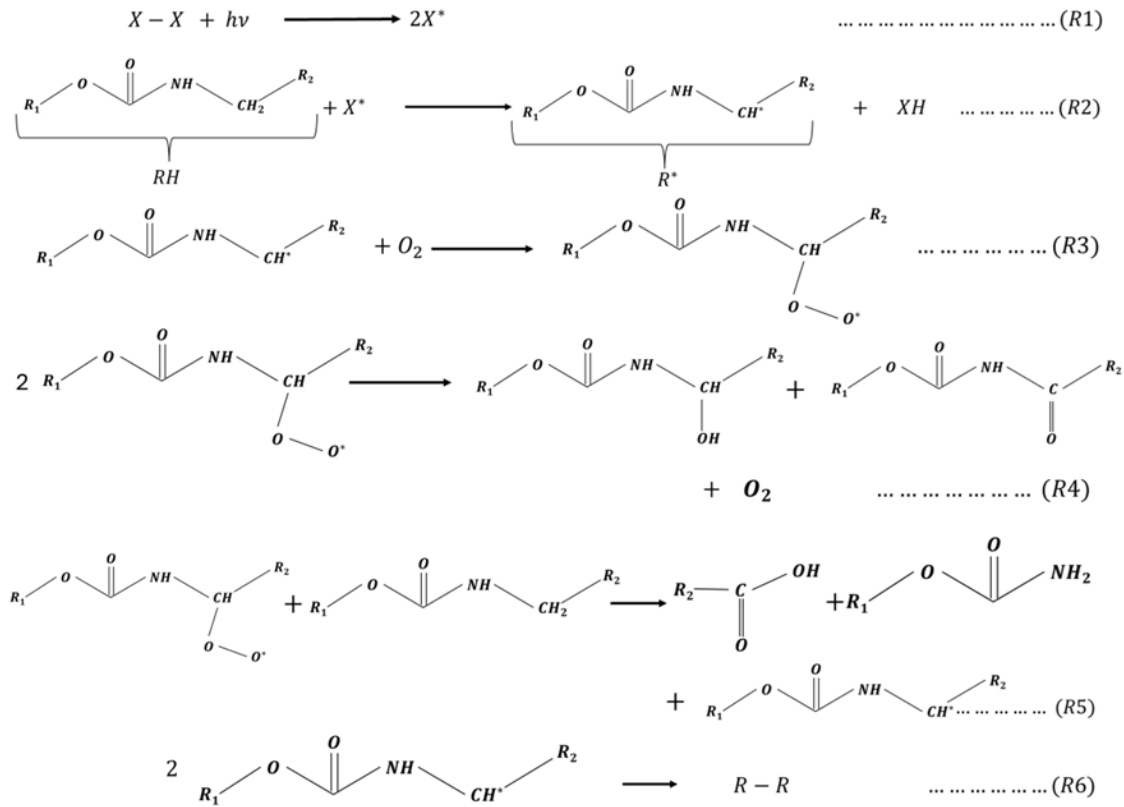

Figure 1. Schematic illustration of photodegradation mechanism of acrylic-polyurethane coating. $R_1$ and $R_2$ symbolize the remaining part of polymer.

Here $X - X$ represents the chromophores (light-absorbing groups) present in the coating matrix, which absorb the UV energy and results in the formation of highly reactive species, called as free radicals. These impurity free radicals (represented by $X^*$), abstracts the hydrogen from polymer (represented by $RH$) within the coating matrix and generates polymer free radical (represented by $R^*$) as shown in reaction R2. The polymer free radical generated initiates a cascade of reactions within the coating chemistry. These reactions can react with the oxygen from the atmosphere, leading to the formation of oxygenated species like peroxy radical in reaction R3. Subsequently. these peroxy radicals can lead to chain scission by breaking down of polymer chains into smaller fragments as seen in reaction R5, reducing the molecular weight of coating, and causing a reduction in mechanical properties like tensile strength and flexibility. While other free radicals may form new chemical bonds between peroxy radicals or polymer chains as observed in reactions R4 and R6, respectively. Crosslinking between polymeric chain leads to increased brittleness and potential cracking of coating. Reactions R1, R2, R3 and R6 are considered elementary, while reactions R4 and R5 are overall reactions. Reactions R4 and R6 are termination reactions.

The degradation reactions in the mechanism mentioned above are fueled by the components within the coating itself, with atmospheric oxygen playing a crucial role. As oxygen diffuses from the surface into the coating's depth, it becomes a key reactant in the photodegradation process. Therefore, diffusion of oxygen is also incorporated into the degradation mechanism along with other reactions.

### B. Mathematical modeling

A transient 1-D model (in depth *y*) capable of describing degradation of a coating exposed to UV radiation and humidity is described here. The model considers photoinitiated chain reactions, oxygen diffusion, and reduction of crosslinking density from the degradation mechanism. The overall purpose of the model is to estimate the mass loss and matrix ablation of the coating, along with the concentration profiles of components like polymer, oxygen, and volatile compounds. The assumptions underlying the model development are:

- Coating is a combination of a clearcoat and topcoat, with no additives, UV absorbers, or antioxidants.
- Coating matrix is ideal and does not contain loops or intramolecular bonds after curing.
- The initial coating surface is flat, and free of any defects.
- The diffusion of active components other than oxygen is assumed to be negligible.
- Coating is isothermal, with its temperature remaining the same as the surrounding environment.

- No thermal degradation occurs in the coating matrix at 300K.
- Exposure conditions mimic those of accelerated weathering conditions, with constant air temperature, relative humidity, and UV radiation intensity.
- Radiation intensity received by the coating corresponds to the radiation intensity emitted by UV lamps in accelerated weathering chambers.

The governing equations of the mathematical model are formulated by incorporating the kinetics of various reactions of degradation. The initial concentrations and boundary conditions for all involved components were derived based on the coating chemistry and testing conditions. These components/species, along with concentration expression and their initial concentrations and boundary conditions used in the model are summarized in Table 1.

*Table 1: Components involved in the degradation reactions along with their concentration expressions, initial concentrations, and boundary conditions.*

| S. No. | Component / Specie | Concentration expression | Initial concentrations and boundary conditions |
|---|---|---|---|
| 1 | Main polymer ($R_1$-COONHCH$_2$-$R_2$) | $[RH]$ | Initial Cross-linking density |
| 2 | Impurity (X-X or X-H) | $[X-X]$ | 1% of initial cross-linking density [21] |
| 3 | Impurity radical (X*) | $[X^*]$ | 0 |
| 4 | Polymer radical (R*) | $[R^*]$ | 0 |
| 5 | Oxygen ($O_2$) | $[O_2]$ | At time, $t=0 \begin{cases} S_{eq} \times P_{O_2} \text{; at top of the coating} \\ 0 \text{; at all other part of coating} \end{cases}$ <br> At all times $\begin{cases} \frac{\partial [O_2]}{\partial y} = 0 \text{; at bottom of the coating} \end{cases}$ |
| 6 | Per-oxy radical | $[ROO^*]$ | 0 |
| 7 | Alcohol | $[ROH]$ | 0 |
| 8 | Ketone | $[R_2CONH_2]$ | 0 |
| 9 | Carboxylic acid | $[R_2COOH]$ | 0 |
| 10 | Urethane product | $[RO]$ | 0 |
| 11 | Crosslinked polymer (R-R) | $[R-R]$ | 0 |

### B.1. Chemical Kinetics

The chemical kinetics (rate of formation or loss) for all the involved components/species is calculated using mass balance and rate laws. A typical mass balance for a component can be written as:

$$Unsteady\ state\ term = Diffusion + Source - Convection \qquad (1)$$

In this study, the mass balance is performed with respect to concentration. The unsteady source term in equation (1) is represented as a partial differential operator, accounting for changes in concentration over both time and space of coating.

Among the various species, only oxygen was considered a diffusing species, making the diffusion term non-zero in the oxygen mass balance. For all other species, the diffusion term was assumed to be zero. The resulting mass balance equation for oxygen is as follows:

$$\frac{\partial [O_2]}{\partial t} = \frac{\partial}{\partial y}\left(D_{O_2}\left(\frac{\partial [O_2]}{\partial y}\right)\right) - k_3[R^*][O_2] + k_4[ROO^*]^2 \qquad (2)$$

Here, the diffusion term is derived using Fick's law of diffusion. In the equation, $y$ denotes the dimension along the depth of coating and $t$ denotes the time of weathering, $k_3$ and $k_4$ are the rate constants of reaction R3 and R4, respectively. While $[O_2]$, $[R^*]$, $[ROO^*]$ represents the concentration of oxygen, polymer radical, and peroxy radical respectively, $D_{O_2}$ is the diffusivity of oxygen in acrylic-polyurethane coating.

Oxygen is assumed to be initially present at the surface of coating, sourced from the atmosphere. While at the bottom of the coating ($depth\ =\ l$), the concentration of oxygen is always assumed zero. The boundary conditions for oxygen are defined as follows.

$$[O_2](t = 0, y) = 0 \qquad (3)$$

$$[O_2](t, y = 0) = S_{eq} \times P_{O_2} \qquad (4)$$

$$\frac{\partial [O_2]}{\partial y}(t, y = l) = 0 \qquad (5)$$

Where, $S_{eq}$ is the solubility of oxygen in polyurethane coating film, $P_{O_2}$ is the partial pressure of the oxygen under environmental conditions.

For other components, the mass balance using the reactions in mechanism resulting in governing equations are presented in Table 2. The initial concentrations and boundary conditions related to each governing equation or component are also summarized. Here, all the symbols used are described in the Nomenclature, while concentration expression for each component is described in Table 1.

*Table 2: Governing equations developed from the mass balance of each species involved in degradation mechanism along with their boundary conditions.*

| Governing equation | | Boundary conditions | |
|---|---|---|---|
| $\dfrac{\partial [RH]}{\partial t} = -k_2[X^*][RH] - k_5[ROO^*][RH]$ | (6) | $[RH](t=0,y) = [RH]_0$ | (7) |
| $\dfrac{\partial [X-X]}{\partial t} = -k_1\varphi[X-X]E_o e^{-\alpha y[RH]} + k_2[X^*][RH]$ | (8) | $[XH](t=0,y) = 0.1[RH]_0$ | (9) |
| $\dfrac{\partial [X^*]}{\partial t} = 2k_1\varphi[X-X]E_o e^{-\alpha y[RH]} - k_2[X^*][RH]$ | (10) | $[X^*](t=0,y) = 0$ | (11) |
| $\dfrac{\partial [R^*]}{\partial t} = k_2[X^*][RH] + k_5[ROO^*][RH] - k_3[R^*][O_2] - k_6[R^*]^2$ | (12) | $[R^*](t=0,y) = 0$ | (13) |
| $\dfrac{\partial [ROO^*]}{\partial t} = k_3[R^*][O_2] - k_4[ROO^*]^2 - k_5[ROO^*][RH]$ | (14) | $[ROO^*](t=0,y) = 0$ | (15) |
| $\dfrac{\partial [ROH]}{\partial t} = k_4[ROO]^2$ | (16) | $[ROH](t=0,y) = 0$ | (17) |
| $\dfrac{\partial [RO]}{\partial t} = k_4[ROO^*]^2$ | (18) | $[RO](t=0,y) = 0$ | (19) |
| $\dfrac{\partial [R_1CONH_2]}{\partial t} = k_5[ROO^*][RH]$ | (20) | $[R_1CONH_2](t=0,y) = 0$ | (21) |
| $\dfrac{\partial [R_2COOH]}{\partial t} = k_5[ROO^*][RH]$ | (22) | $[R_2COOH](t=0,y) = 0$ | (23) |
| $\dfrac{\partial [R-R]}{\partial t} = 0.5\, k_6[R^*]^2$ | (24) | $[R-R](t=0,y) = 0$ | (25) |

### B.2. Kinetics solution procedure

The developed kinetics model was solved using a combination of finite difference numerical techniques. The governing equations in the model consist of both ordinary and partial differential equations. The ordinary differential equations, which involve time differentials, were solved using the 4th-order Runge-Kutta method (RK4). Meanwhile, the partial differential equation, which contains differential terms in both space and time, was addressed using the Control Volume method. This approach linearized the equation into time differentials across space for efficient numerical solution.

For numerical solving, the 1-D coating of 70 µm depth discretized into space of 1 µm grid length was taken in this model, as shown in figure 2. The timestep was taken as 0.5s to maintain the stability of solution using finite difference methods. Total simulation time for the model encompasses the timestep for solving and total time for which weathering / degradation was studied.

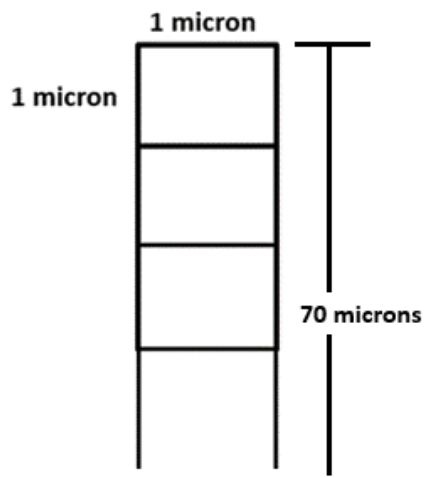

*Figure 2. Schematics for the discretized coating along the depth of coating.*

### B.3. Estimation of model parameters

The model requires a number of physical and chemical constants or parameters, which were taken from the coating chemistry and environmental conditions. However, certain system parameters were not available from the literature, which were estimated using a systematic optimization procedure. The known parameters (from literature) and initially assumed parameters for the kinetics model are represented in Table 3.

*Table 3: Initial system parameters for the chemical kinetics model.*

| Parameter with SI Units | Values |
|---|---|
| $[RH]_0 \, (mol/m3 \, of \, coating)$ | 4082 [18] |
| $P_{O_2} \, (Pa)$ | 21278 [14] |
| $S_{eq} \, (mol/m^3 \, coating \, Pa)$ | $1.6 \times 10^{-4}$ [14] |
| $D_{O_2} \, (m^2/s)$ | $7.5 \times 10^{-13}$ [22] |
| $\varphi \, (dimensionless)$ | $4 \times 10^{-6}$ *(assumed)* |
| $E_0 \, (W/m^2)$ | 30 [14-15,20] |
| $\alpha$ | 1 [14] |
| $k_1 \quad (m^2/W.s)$ | 1 *(assumed)* |
| $k_2 \quad (m^3/mol.s)$ | 0.1 *(assumed)* |
| $k_3 \quad (m^3/mol.s)$ | 10 *(assumed)* |
| $k_4 \quad (m^3/mol.s)$ | 0.1 *(assumed)* |
| $k_5 \quad (m^3/mol.s)$ | 0.1 *(assumed)* |
| $k_6 \quad (m^3/mol.s)$ | 1 *(assumed)* |

To estimate the unknown parameters, the model's output product component concentrations were compared to experimental data of coating under similar testing environments [18]. Initially, the kinetics model was solved with an approximate set of system parameters. The total time of degradation was set to 80 days of accelerated weathering time, as in the experimental testing. For concentration matching/comparison of the reactant (polymer) and reaction product (carboxylic acid), experimental data is available in terms of absorbance. The concentration data from the kinetics model was converted into absorbance by integrating the concentration values over the depth to which FTIR waves can penetrate. Absorbance of a component is calculated using the concentration data as follows:

$$A = K \int_{y=0}^{y=70} c(y) \, dy \tag{26}$$

Where, $A$ is the absorbance of the component/specie, $K$ is a constant, and $c(y)$ is the concentration of the component as a function of the depth from 0 µm (surface) to 70 µm (depth up to which FTIR wavers penetrate). Additionally, the absorbance from experimental data and the integrated concentrations from the model solution for each species were normalized to their respective maximum values. These normalized values were then compared to assess the match between the experimental and model data, using both profile curves and mean-squared error (MSE).

To minimize MSE, the approximated system parameters were optimized using multi-objective multi-phased particle swarm optimization algorithm [23]. This algorithm iteratively adjusts the parameters to minimize the discrepancies between the model and experimentally observed concentrations, ensuring a closer alignment with real-world behavior. To improve the optimization times, the partial differential equation based on oxygen diffusion in the model was removed and the average oxygen concentration from the previous iteration was taken. Also, the average oxygen concentration from the previous iteration was scaled appropriately. Thus, the optimization used only the set of ordinary differential equations to estimate the unknown set of system parameters. The multi-phased optimization procedure is represented in figure 3. The optimization was carried out till the error or MSE in the integrated concentration or absorbance fell to the desired value (<0.005).

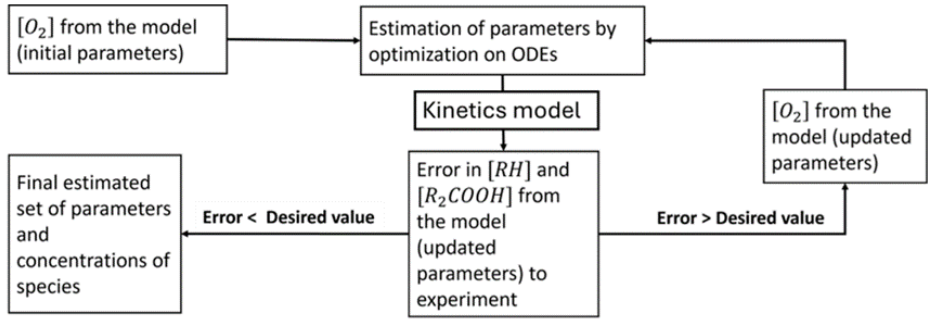

*Figure 3. Flow chart describing the multi-phase optimization procedure to estimate the system parameters of kinetics model.*

Using the estimated set of system parameters from the PSO method, the concentration of chemical species during the accelerated degradation were obtained from the kinetics model. The service lifetime of coating can be determined by the time coating takes place to reach 10% of the concentration of polymer left. At this concentration, the coating is assumed to be ablated and not sufficient to be functional [24].

Additionally, the number of photons absorbed by the coating over time during degradation was recorded. This was done by estimating the total amount of energy from UVA radiation absorbed by the coating. Each photon was assumed to have a wavelength of 340nm, corresponding to a typical range of UV-A radiation. The number of absorbed photons was calculated as the ratio of total energy absorbed to the energy of single photon, using the following equation:

$$N_p = \frac{\int_0^t \int_0^y E_0 \varphi \exp(-\alpha y\, [RH])\, dy\, dt}{h\nu_{UV-A}} \qquad (27)$$

Here, $\nu_{UV-A}$ is the frequency of the photons of UV-A radiation corresponding to 340nm wavelengths.

### C. Modeling of physical degradation

One of the key mechanisms underlying the physical degradation of coatings is the formation and growth of defects, such as cracks, pores, and delamination, within the coating structure. Physical model employs Monte Carlo technique, which simulates these structural changes as surface erosion. Using these evolved surface topography and composition, measurable properties like roughness, thickness loss, mass loss, and related mechanical properties like gloss loss, fracture toughness, etc. are predicted.

With the degradation, the system evolved in which final state was derived by the repeated application of the many small events. The Monte Carlo method assumes the photon flux from UV radiation as the initiator to the photodegradation and each small event is assumed to be a photon from radiation being targeted on the coating. Time required for such an event was represented as the Monte Carlo timestep. The model divides the coating layer into small cells of dimensions representing a minimum volume of coating being ablated by a single photon. This was determined in our work as the number of photons being absorbed by a coating of specific dimension to get ablated.

For an acrylic based polyurethane system, irradiated by UV radiation of 340 nm wavelengths, the average depth of absorption was obtained to be 6.36 micron. The average depth of absorption implies that the majority of the photons (~60%) are absorbed within this depth of the coating. Thus, a coating of dimension $6\mu m \times 1m \times 1m$ (also referred as Representative Volume Element, RVE) was fed to the optimized kinetics model and number of photons required to ablate was calculated. Using these number of photons (~ $5.59 \times 10^{16}$), the minimum volume being ablated by a photon was obtained to be $1.07 \times 10^5 \ nm^3$. Such minimum volume corresponds to a cell of dimension $47.5 \ nm \times 47.5 \ nm \times 47.5 \ nm$. This volume contains sufficient material to be ablated when a reactive photon caused ablation/ photodegradation reaction event occurs and termed as pixel volume for coating damage for further references in this study.

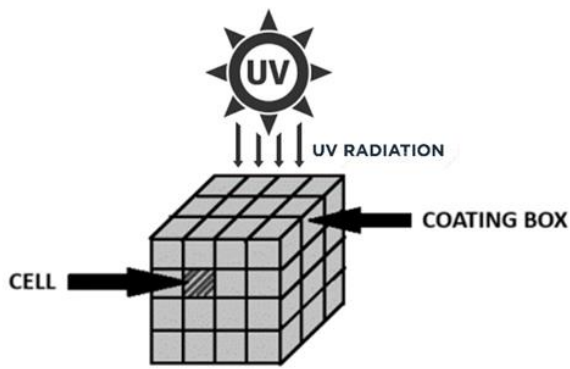

*Figure 4: Schematic illustration of coating divided into cells for simulation in Monte Carlo model.*

The coating divided into small boxes of the determined cell dimension with polymer was used for Monte Carlo simulation as shown in figure 4. During each Monte Carlo (MC) event, a photon was targeted on the random cell of the coating. The developed MC code chose the location of the random cell. As per the location of the cell, the probability ($p$) of the material to absorb incoming photon in the chosen cell was calculated. The code then generated another random number. If this random number was less than the calculated probability of absorption ($p$), then the materials were removed or modified (changed to another material), otherwise not. Every successful and unsuccessful event to ablate material was recorded to identify the timeline of the degradation events.

Based on the location of the degradation event, cells with degraded polymers were separately identified. The cells which were degraded on the surface of the coating escape out and form a vacuum. Such vacuum cells were termed as 'pits', while other types of damaged cells in bulk were termed as 'pore'. For pore cells, the age of the degradation event was recorded. The information of each cell whether it is a pit, pore, or undegraded polymer was represented by a pixel spatiotemporal damage status. After every successful degradation event, pixel spatiotemporal damage status was updated accordingly.

Pores were allowed to diffuse out after they encounter pits with a 4-way connectivity, and then the pores were converted to be a part of the pits. Pores diffuse according to their age. The pores which were generated first (higher age of pore) diffuse out first compared to the pores formed later, even if all were in contact with same pit at same time. This allows the model to mimic the real paint degradation process effectively. As these pores degraded, vacuums were generated at their location and thus the pit grew to the pore location.

During degradation, islands of pores or undegraded polymer were formed which were surrounded by the vacuum and disconnected from the bulk of the coating material. Such islands were termed as unconnected islands and are lost from the coating material during real weathering testing due to rain or wind. An image processing tool like OpenCV in python was utilized to identify these unconnected islands of polymers in our model. The connected component method of OpenCV scanned through the whole coating MC simulation box and segregated out the unconnected islands. The connected component utilized the 4-way connectivity check in the coating and identified the disconnected components. Such unconnected islands were removed to mimic the physical degradation of coating to a closer practical sense. Evolution of coating surface was developed over time in terms of Monte Carlo timestep unit (represented as *MC timesteps*).

### D. Time scaling of physical degradation model

To bridge the gap between Monte Carlo model and calibrated kinetics model, we established a direct relation between Monte Carlo timestep unit and accelerated weathering time. This was achieved meticulously by correlating the coating's polymer concentration predicted by our Monte Carlo degradation model with that observed in kinetic models or experiments. By determining the average ratio of Monte Carlo timesteps to corresponding times in the kinetic models at equivalent polymer concentrations, we effectively scaled the simulation time to real-world weathering. This time-scaling relationship enabled us to express the calculated surface properties, derived from the Monte Carlo simulations, in terms of accelerated weathering time, providing tangible and impactful insights into the coating's long-term performance.

### E. Surface dominated properties

Surface properties play a critical role in the performance and durability of coatings, as changes in chemistry and topography can significantly impact their macroscopic behavior. For instance, thickness loss and surface roughness profiles are directly influenced by the degradation-induced alterations on the coating's surface. These changes in surface roughness, in turn, govern crucial properties like gloss and wetting contact angle. Furthermore, the presence of surface defects can act as initiation sites for cracks, ultimately dictating the coating's fracture toughness. Using the chemistry and topographical changes evolved during the Monte Carlo simulations, these surface dominated properties were predicted using empirical relations.

## 3. Results and discussion

### A. Estimated kinetics model parameters

Before the optimization procedures, the kinetics model using assumed system parameters predicted concentration profiles, which exhibited significant deviations from the experimental data, leading to higher-than-desired MSE values. Through iterative optimization, the MSE was progressively reduced, eventually dropping below the desired threshold for both components as mentioned in Appendix. During the optimization iterations, the time-series scaling of oxygen concentration was carried out to achieve faster optimization. These scaled bulk-averaged concentration profiles of the oxygen for certain iterations sequences are mentioned in Appendix.

Table 4: Optimized set of unknown parameters for the chemical kinetics model.

| Parameter with SI Units | Values |
|---|---|
| $\varphi \ (dimensionless)$ | $4.18 \times 10^{-6}$ |
| $k_1 \quad (m^2/W.s)$ | $1.52 \times 10^{-4}$ |
| $k_2 \quad (m^3/mol.s)$ | 12.1 |
| $k_3 \quad (m^3/mol.s)$ | 1.03 |
| $k_4 \quad (m^3/mol.s)$ | 3.09 |
| $k_5 \quad (m^3/mol.s)$ | 4.42 |
| $k_6 \quad (m^3/mol.s)$ | 2.33 |

The unknown parameters of the kinetics model were successfully estimated and are presented in Table 4. With these optimized system parameters, the kinetics model accurately predicted the concentration profiles of the components, as illustrated in figure 5. The strong agreement between the predicted profiles and the available experimental data provides successful validation of the kinetics model. Additionally, the change in mass of a remaining component of a polymer ($R_1$ or $R_2$) in the coating has been observed to be very small over the degradation, as shown in figure 6. It shows the correctness of the reaction mechanism and parameters used in the model.

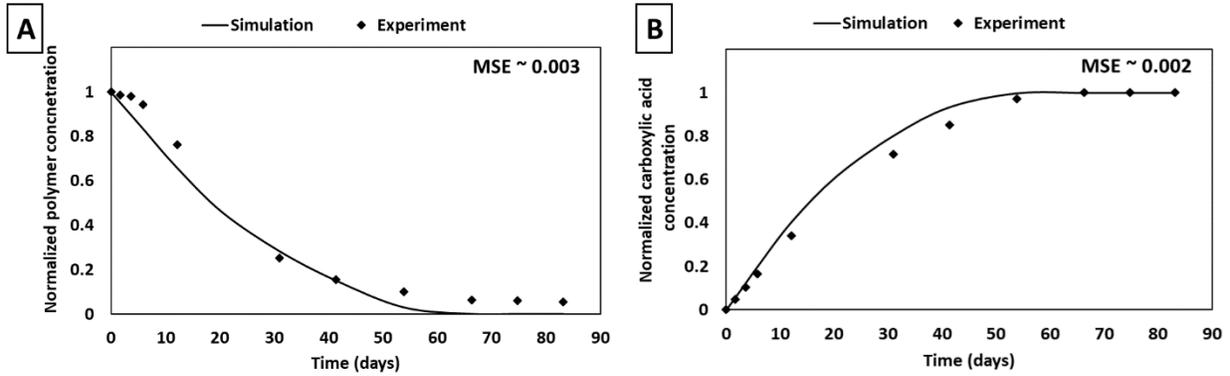

*Figure 5: Comparison between experiments (diamonds) [18] and simulations results (solid lines) from kinetics model. (A) and (B) represent the normalized polymer and normalized carboxylic acid concentrations, respectively.*

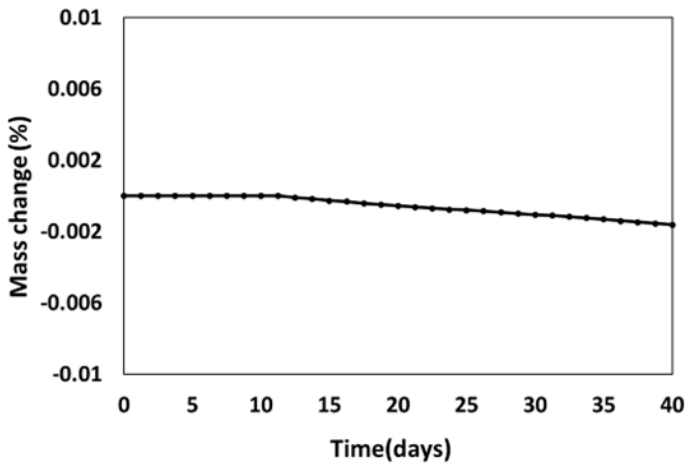

*Figure 6: Percentage change in mass of polymeric components ($R_1$ or $R_2$) in the coating.*

### B. Concentration profiles from kinetics model

Using the optimized set of kinetics parameters, the concentration profiles of the chemical species (polymer, carboxylic acid, oxygen) with time at different depths of coating were obtained as shown in figure 7. The polymer concentration at the surface of the coating decreases at a faster rate as compared to at other depths as shown in figure 7(B). This difference in rate is maybe attributed to the higher amounts of UV radiation and oxygen concentrations at the surface. Similarly, products such as carboxylic acid were observed at the surface and later in the depths of the coating during degradation as shown in figure 7(C).

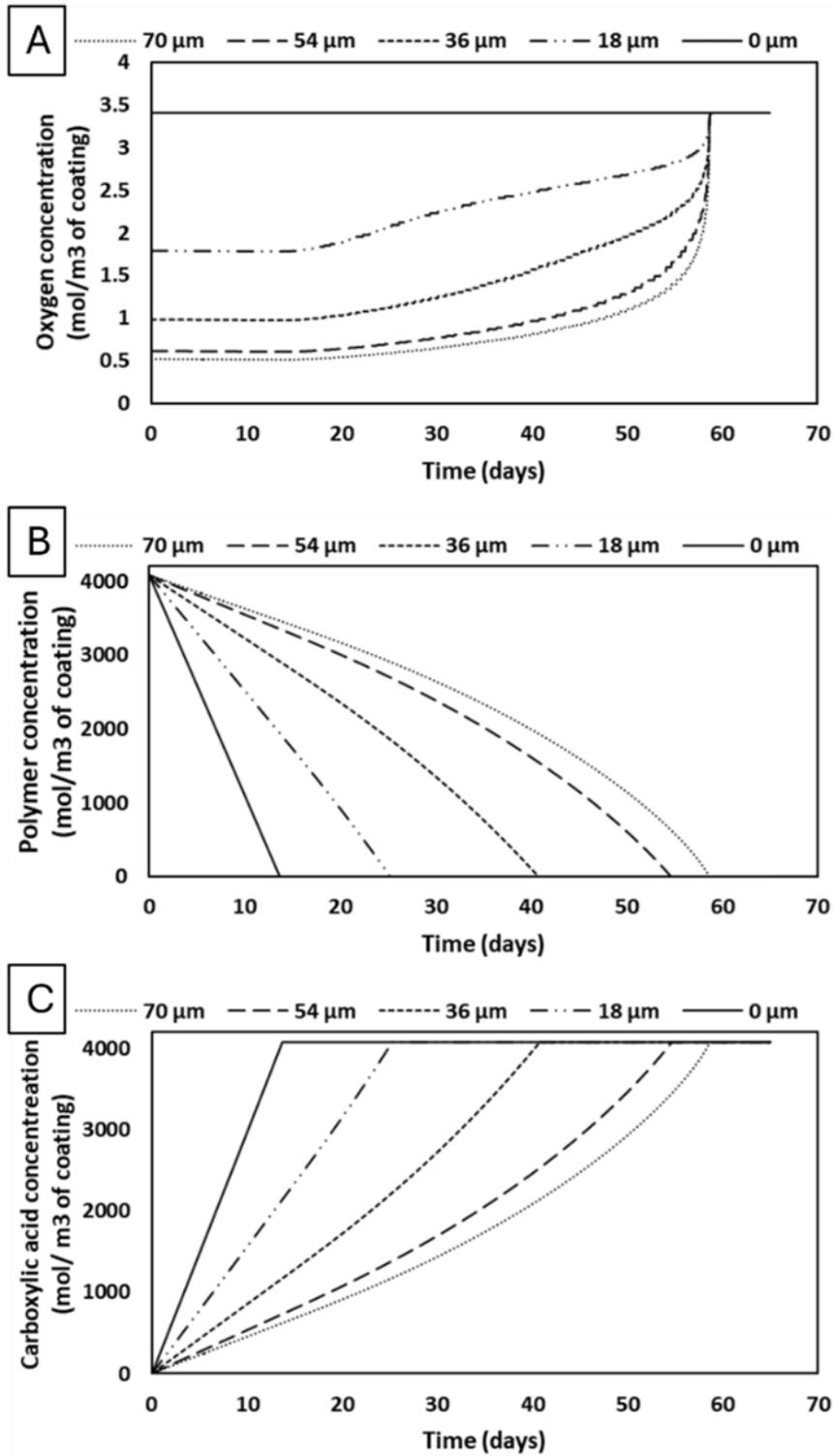

*Figure 7: Depth-wise concentration profiles of components (A) Oxygen, (B) Polymer, and (C) Carboxylic acid in the coating during degradation from kinetics model.*

## C. Surface topography of coating

Figure 8 represents the evolution of the coating surface over time from the Monte Carlo model. In this degradation simulation, the unconnected islands of polymer/ debris were not removed. As the material degraded or eroded, pits started to appear and grew in the depth of the coating. A similar trend is observed in the case of simulation when the unconnected islands of polymer/ debris were removed as shown in figure 9. The islands of polymer (degraded or undegraded) encircled and highlighted in figure 9 (A) were identified successfully and removed in figure 9(B).

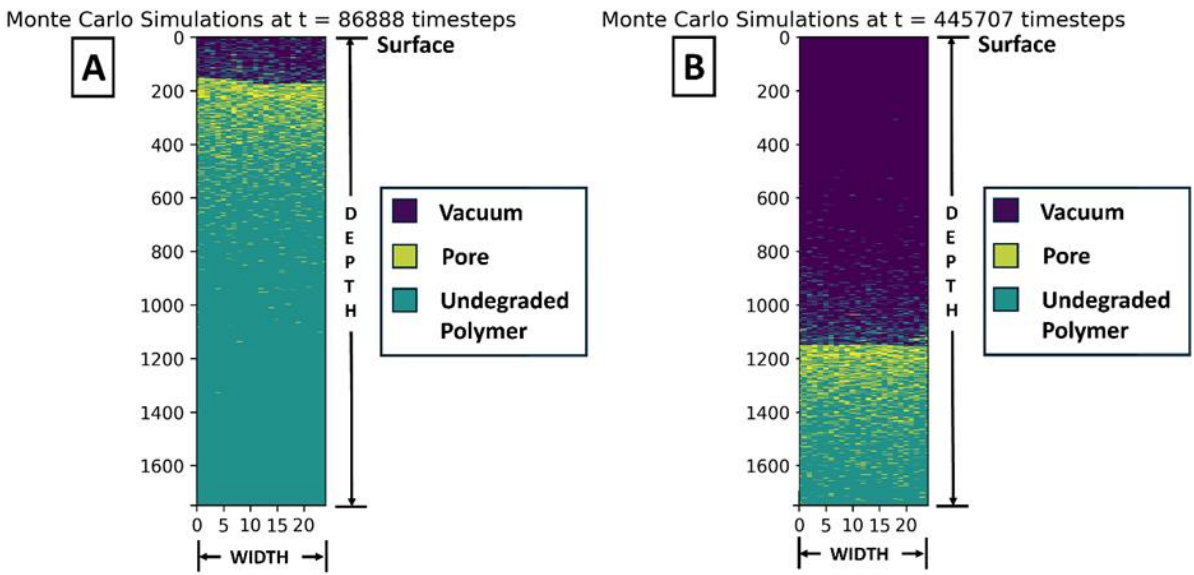

*Figure 8: Surface topography along the depth of a coating (40 nm x 1 µm x 70 µm) from Monte Carlo model at (A) t = 86888 MC timesteps and (B) t = 445707 MC timesteps, respectively.*

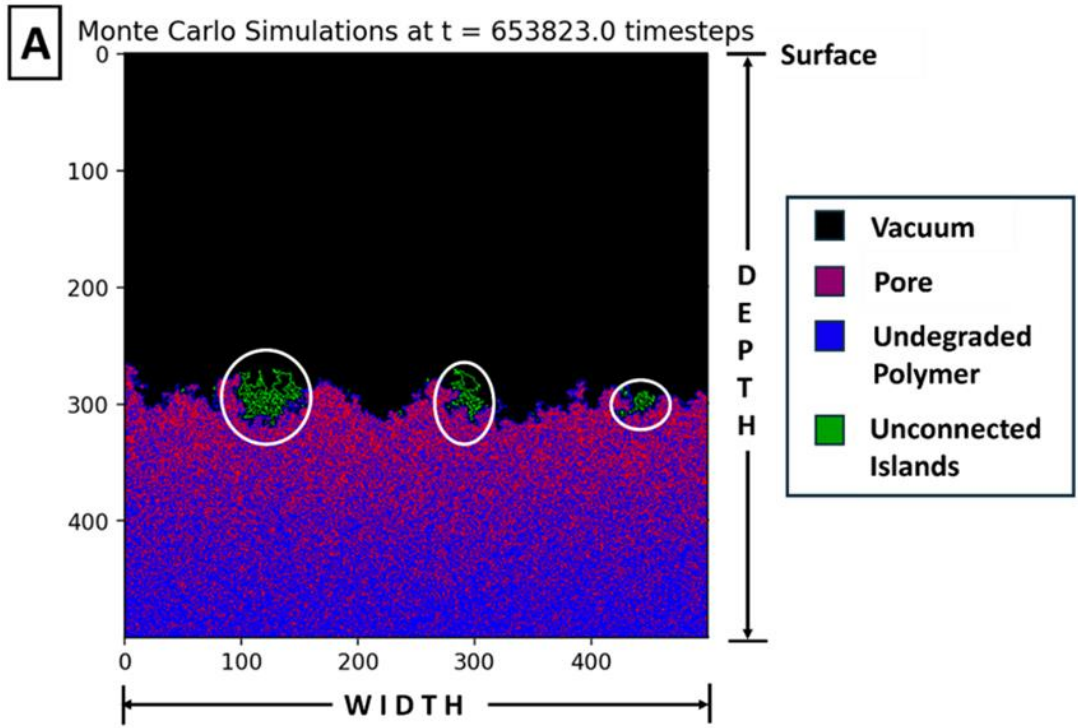

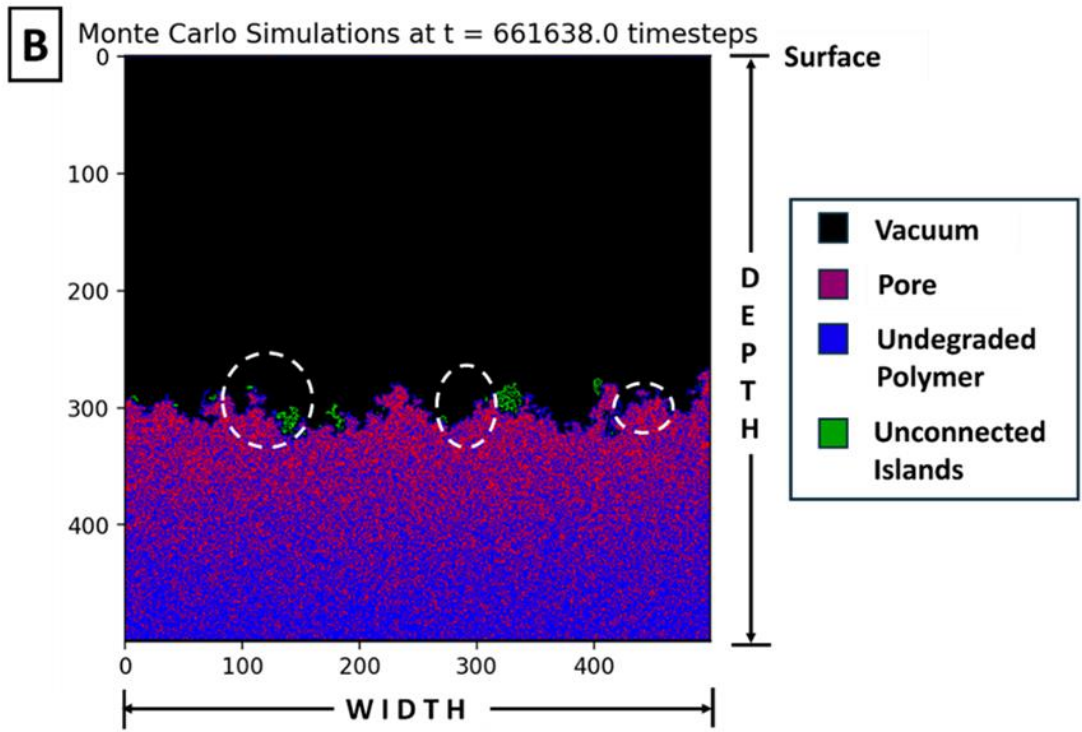

*Figure 9: Surface topography along depth of coating (40 nm x 20 μm x 20 μm) from Monte Carlo model at (A) At t = 653823 MC timesteps, unconnected islands are identified (green color) and encircled, (B) At t = 661638 MC timesteps, unconnected islands identified earlier are removed as shown by the dotted circles at same place as before.*

### D. Scaled timestep of Monte Carlo

To bridge the gap between Monte Carlo simulations and real-world weathering, a time scaling factor was established. This factor translates the simulation's timesteps into accelerated weathering time, enabling a direct comparison with experimental degradation kinetics.

Two distinct scenarios were investigated: one where unconnected island in the simulation were retained, and another where they were removed. Notably, the time scaling factor differed significantly between the two cases: 5.39 seconds for the former and 1.77 seconds for the latter, highlighting the influence of unconnected islands on the degradation process. This time scaling proved highly effective in aligning the Monte Carlo simulations with the established kinetics model. Figures 10 and 11 illustrate the remarkable agreement between the simulated and modeled concentration profiles of the polymer, both with and without the removal of unconnected islands. This successful correlation underscores the validity of the time scaling approach, enabling researchers to draw meaningful inferences about real-world degradation behavior directly from the Monte Carlo simulations.

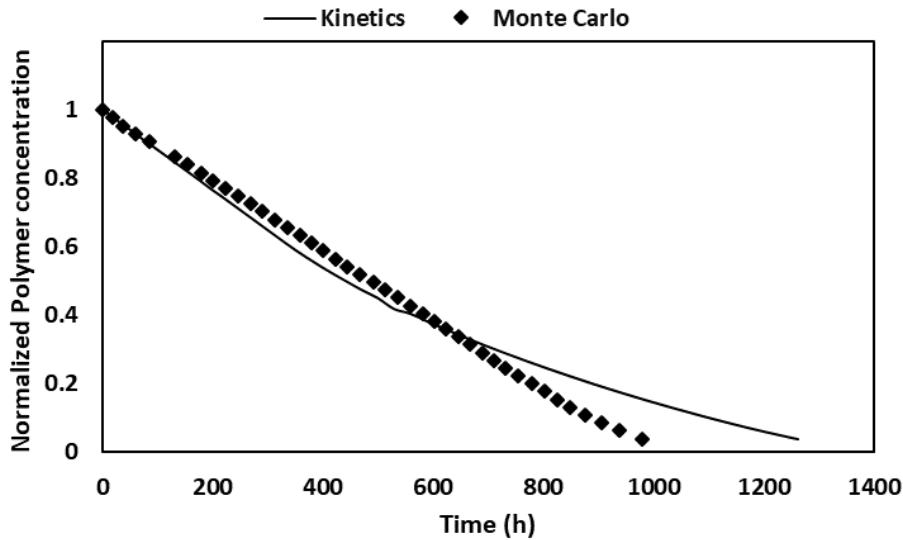

*Figure 10: Comparison of the normalized polymer concentration between chemical kinetics model and Monte Carlo model for a 70-micron depth coating for accelerated weathering. In Monte Carlo simulations, the unconnected islands were not removed.*

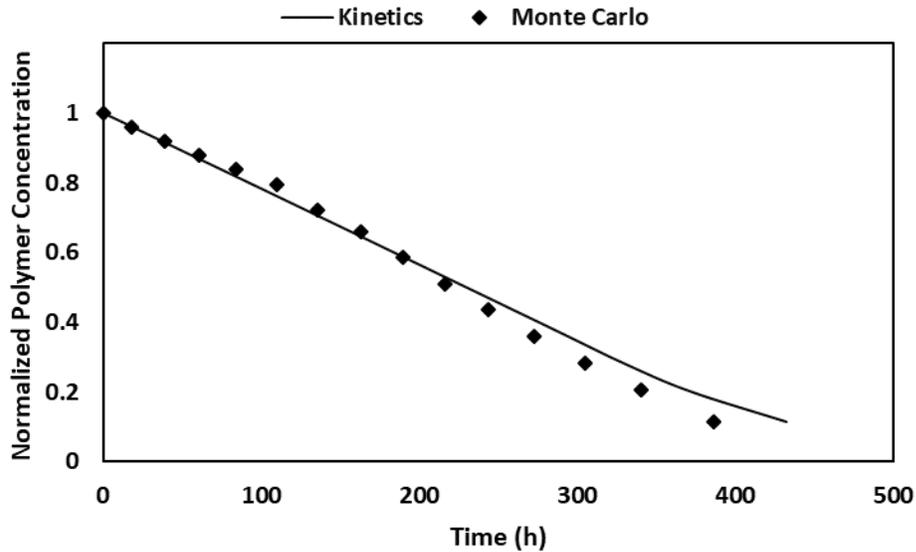

*Figure 11: Comparison of the normalized polymer concentration between chemical kinetics model and Monte Carlo model for a 20-micron depth coating. In this Monte Carlo simulation, the unconnected islands were removed.*

### E. Surface dominated properties

Using the time scaling factor and surface topography obtained from the Monte Carlo model, surface dominated properties were calculated in accelerated weathering time.

### E.1. Thickness loss

From the concentration profiles of polymer in the coating, the loss in thickness was evaluated by depth up to which the polymer has been removed on average. Thickness loss for a 70-micron depth coating sample with respect to accelerated weathering time is shown in figure 12(A). A similar trend in the decrement of the coating thickness has been obtained experimentally for an epoxy based polyurethane coating sample under natural weathering exposure as shown in figure 12(B) [25].

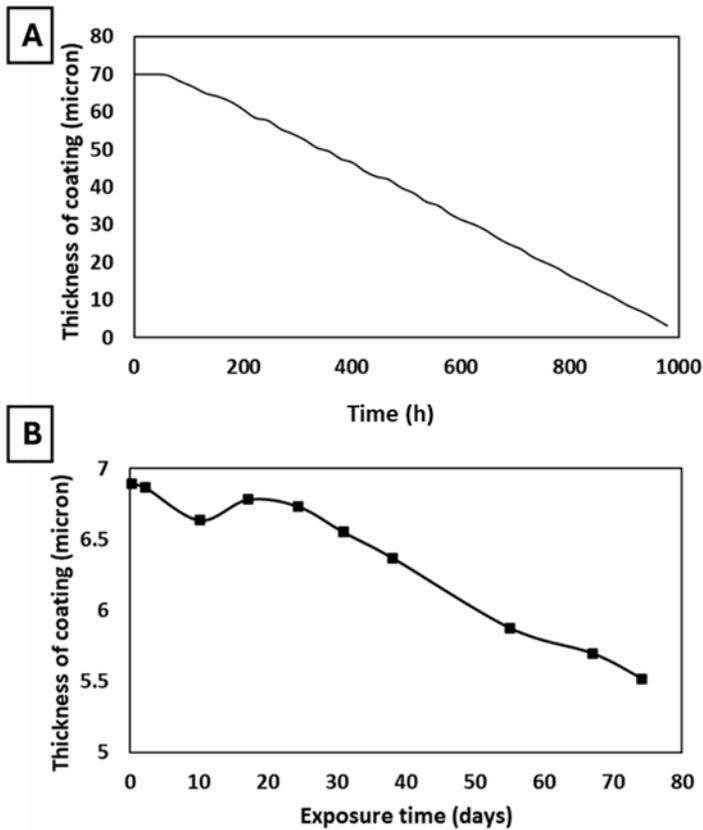

*Figure 12: Thickness of the coating varying with time depicting the loss (A) with accelerated weathering time from Monte Carlo model, (B) with natural outdoor exposure of epoxy-based PU coating experimentally [25].*

### E.2. Surface profiles

As the coating degrades, the thickness of coating starts to decrease due to the loss of polymer from the surface. This polymer loss is not uniform throughout the surface and results in roughening of the surface. From the Monte Carlo simulation, the change in polymer concentration at the surface up to 40 nm resolutions were obtained. Scans of varied sizes (40nm x 40nm or 200nm x 200nm or 1000nm x 1000nm) can be made and depth of coating over those scans is obtained. Using such scans of 40nm x 40nm, the evolution of surface profiles was obtained as shown in figure 13.

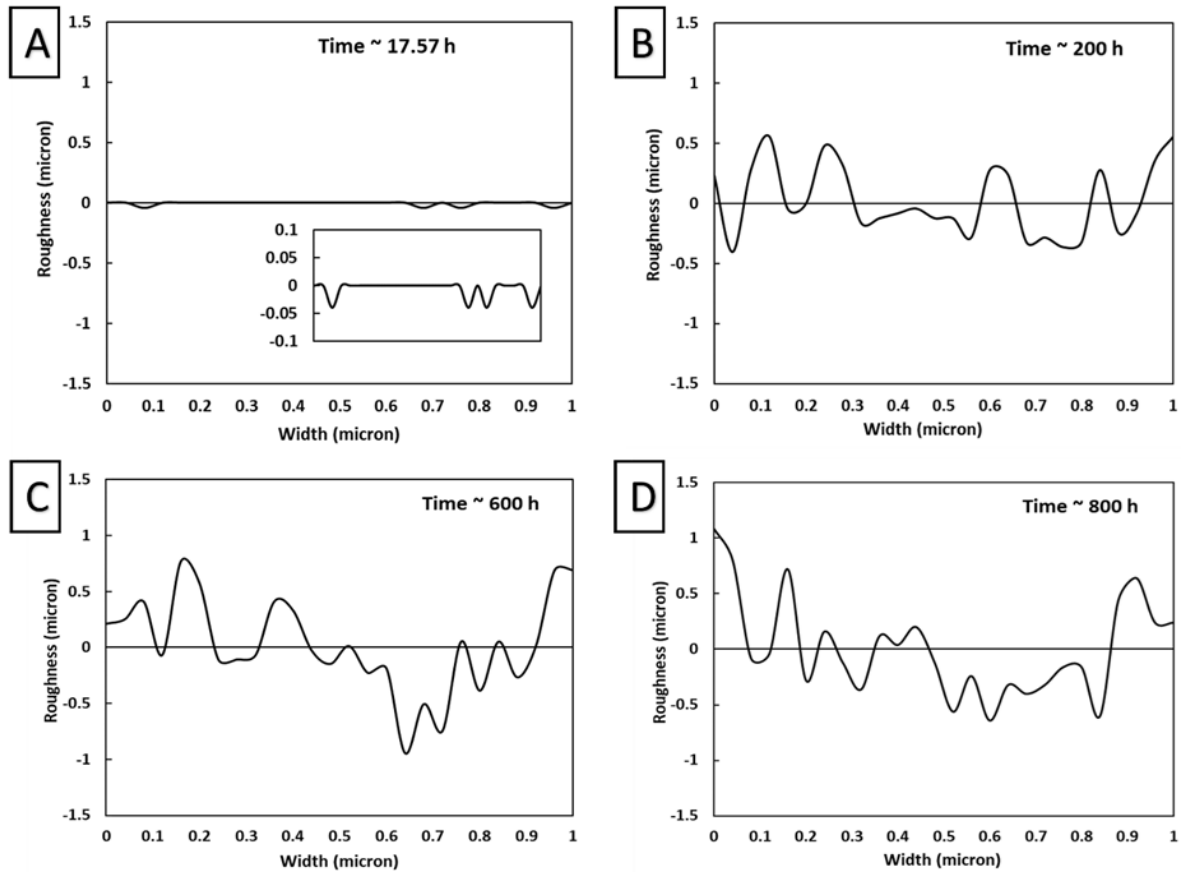

*Figure 13: Surface roughness profiles of a 70-micron depth coating developed from the Monte Carlo model at various times of (A) 17.57 h with zoomed profile (inset), (B) 200 h, (C) 400 h, (D) 800 h.*

**E.3. Surface roughness parameters**

Using the above obtained surface profiles, RMS roughness was calculated by the standard deviation of the roughness profile data. RMS roughness for the 70-micron coating from Monte Carlo model is shown in figure 14. While precise RMS roughness data for our specific polymer coating under the modeled degradation conditions were unavailable, Figure 15 presents the roughness variation observed in an acrylic-based polyurethane system under different exposure conditions [26]. This data serves as a valuable reference point for comparison. RMS roughness can be divided into two different regions according to their characteristics. Region I in figure 14 shows an exponential increase in RMS roughness during initial exposure. This exponential increment trend has been observed experimentally for similar types of coating under different exposure as shown in figure 15. Region II marks the saturation of RMS roughness by following a parabolical trajectory in figure 14. This trend is observed after a certain time of weathering exposure has been carried. This trend is also observed in figure 16 at the end of exposure times.

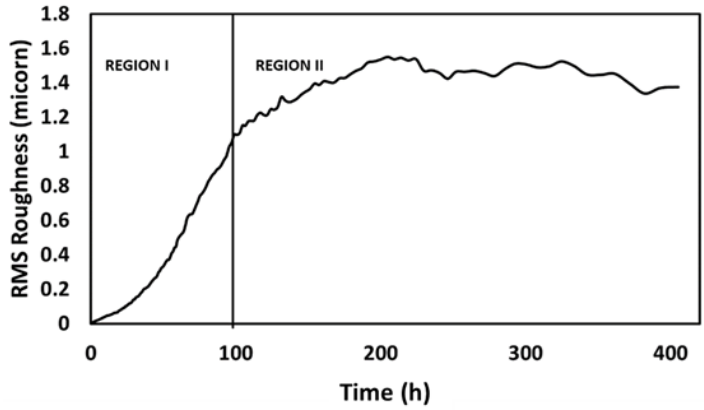

Figure 14: RMS Roughness of the 70-micron depth coating with accelerated weathering degradation time evaluated from the Monte Carlo model.

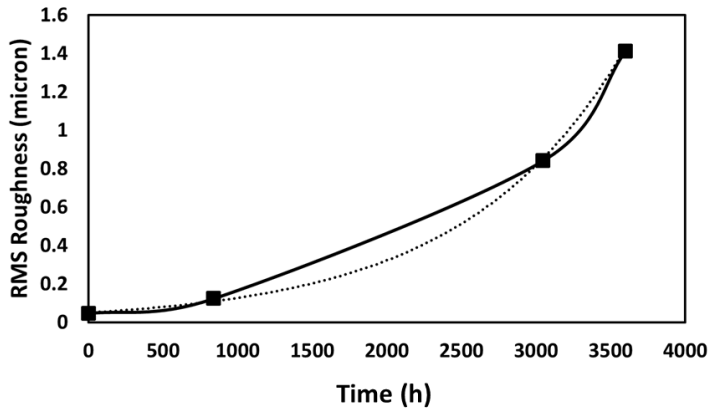

Figure 15: RMS Roughness of the acrylic based polyurethane coating from weathering under different accelerated exposure conditions [26]. These qualitatively verify the RMS roughness obtained from our Monte Carlo model of coating degradation.

Additionally, to further validate our model, the predicted RMS roughness range was compared against existing literature on similar polymer coatings subjected to comparable degradation exposures. Average roughness was calculated corresponding to the mathematical formulation described in ISO 4287 (1997)-surface texture as follows [27].

$$R_a = \frac{1}{w} \int_0^w |h(x)|\, dx \qquad (28)$$

Here, $w$ is the width of the coating and $h(x)$ denotes the height of the coating at the '$x$' distance over the coating's width.

For acrylic based polyurethane coating, RMS roughness values under natural outdoor exposure has been reported to increase to 0.987 µm from 0.199 µm [28]. In another literature for coating under natural exposure, the RMS roughness has been reported to be 1.39 µm after 5 years [21]. RMS roughness for coating under accelerated degradation from our combined model predicted the variation in same range (0 – 1.5 µm).

Other roughness parameters like $R_a$ ($average\ roughness$), $R_z$ and $R_t$ can also be calculated. Average roughness for acrylic based polyurethane coating has been reported to vary in the range of 0.134 µm to 0.580 µm under natural exposure [28]. Similar range (0 to 0.35 µm) of average roughness was obtained from our combined Physical Chemical weathering model.

### E.4. Fracture toughness

A failure criterion for a coating may be crack propagation through the coating material causing channeling cracks, delamination etc. By applying the Griffith crack criterion and using the maximum flaw size generated, the relative strength of the coating with time can be estimated. The maximum flaw size in the surface of coating was calculated from the surface profiles. By Griffith criterion [29], fracture toughness was evaluated as

$$\sigma_G = \sqrt{\left(\frac{2E\gamma_s}{\pi a}\right)} \qquad (29)$$

Where, $\sigma_G$ is the critical stress at which crack propagates, $E$ is the Young's modulus, $\gamma_s$ is the surface free energy per unit area, and $a$ is the maximum flaw size. The fracture strength is represented as a function of its initial value, which eliminates the need-to-know other parameters (Young's modulus and surface energy per unit area) in the Griffith failure criterion. Flaw size for crack is incorporated as a change in surface roughness. The coating fails when the flaws increase to an extent which generates a stress state which it can withstand.

At each time step, the model scans the surface for the deepest pit that has been developed in the coating. This depth was taken as the flaw size to be used in the Griffith equation. Thus, with the evolution of surface in Monte Carlo simulation, the progress of fracture toughness is monitored in sense of relative to initial. The fracture toughness of the coating shows an early accelerated decrease and then a more gradual decrease as shown in figure 16. A similar trend of decrement in fracture toughness has been observed in published experimental literature for a smooth initial surface [30,31].

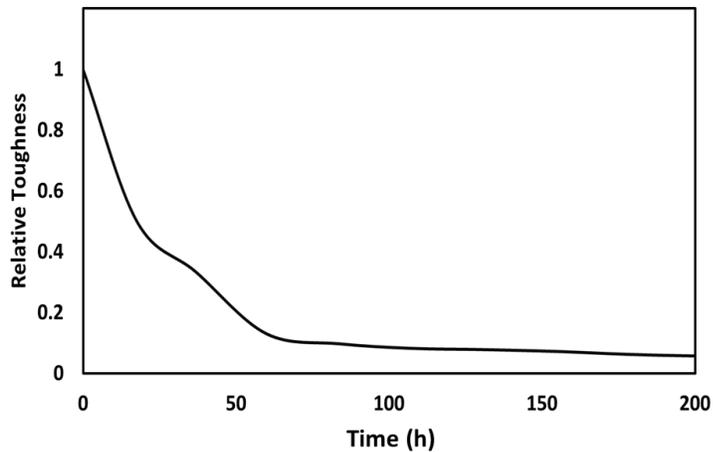

Figure 16: Change in relative fracture toughness of the coating with time of degradation from Monte Carlo model.

Service life of the paint / coating can be calculated by the estimation of the above calculated properties as

$$Service\ life = Total\ life\ of\ coating\ as\ per\ the\ critical\ value\ of\ property -$$
$$Actual\ life\ of\ coating\ passed\ as\ per\ property\ current\ value$$

## 4. Conclusions

An efficient kinetics model has been developed which describes coating degradation under accelerated artificial exposure conditions. The photodegradation mechanism, oxygen permeability and diffusion, reduction of cross-linking polymer density were considered to determine the chemical modifications in the coating system. Experimental data available for the acrylic-polyurethane was used to optimize the model parameters. The concentration profiles of the species involved in the polymer degradation were obtained successfully using the kinetics model. Polymer was observed to degrade at a faster rate at the surface of the coating as compared to in the bulk.

The Monte Carlo model was developed to predict physical degradation in the surface of coating. The changes in the properties of the system were predicted qualitatively based on the theoretical and empirical relations. Based on these measurable properties like thickness loss, roughness parameters, relative fracture toughness, gloss loss, and wetting contact angle, service life of the coating system can be estimated. RMS roughness of the coating surface showed an exponential increase initially and then saturated off with a parabolic increment with exposure time. All the measurable properties obtained showed a similar trend with exposure time as observed in the experimental investigations.

A correlation was established between the accelerated weathering chemical kinetics and physical Monte Carlo model successfully. The correlation enabled the simultaneous prediction of both physical and chemical changes of the coating systems for significantly more accurate service life predictions. This combined model approach offers a more holistic understanding of degradation behavior as compared to the isolated chemical or physical models, which could predict only one of the chemical or physical changes in the coating. Additionally, limitation of predicting the physical changes in terms of physical model arbitrary unit step was eliminated using this combined model approach. The combined model allows for multi-factor failure modes to be studied for coating. With more detailed physical and chemical degradation mechanisms, optimized maintenance and repair schedules can be planned effectively. In summary, combining physical and chemical models for coating degradation results in more robust, accurate, and realistic assessments of material performance, helping engineers and researchers to design better coatings, optimize maintenance strategies, and extend the lifespan of the materials.

Furthermore, this combined model can be employed to explore the impact of additives and stabilizers on coating properties, enabling researchers to optimize formulations for enhanced durability and performance, ultimately revolutionizing coating design by accurately predicting long-term performance and enabling the optimization of protective formulations for more durable and resilient coatings.

## Appendix: Concentration profiles from Optimization Iterations

The comparison of concentration profiles between experiments and predicted using the kinetics model before and during the optimization procedure are as shown in figure A1 and A2 respectively.

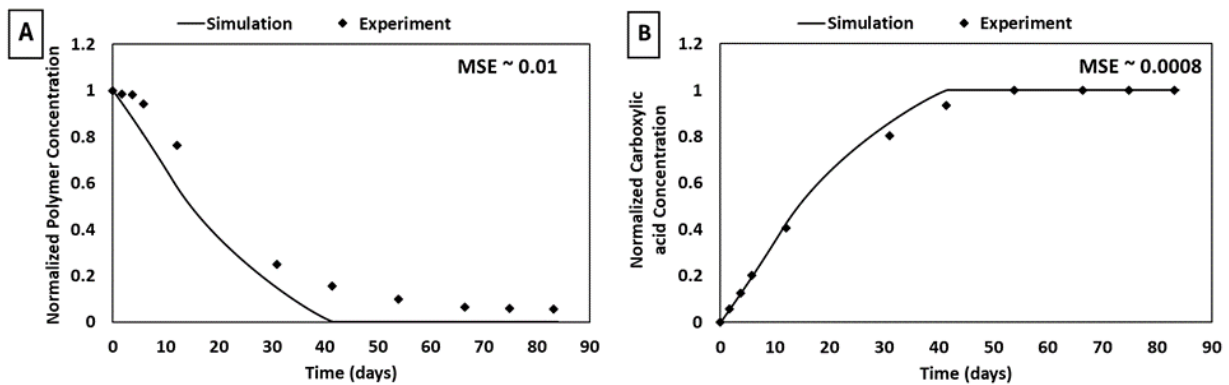

*Figure A1: Comparison between experiments (diamonds) [18] and simulations results (solid lines) from kinetics model using initialized system parameters before optimization procedure. (A) and (B) represents the normalized polymer and normalized carboxylic acid concentrations, respectively.*

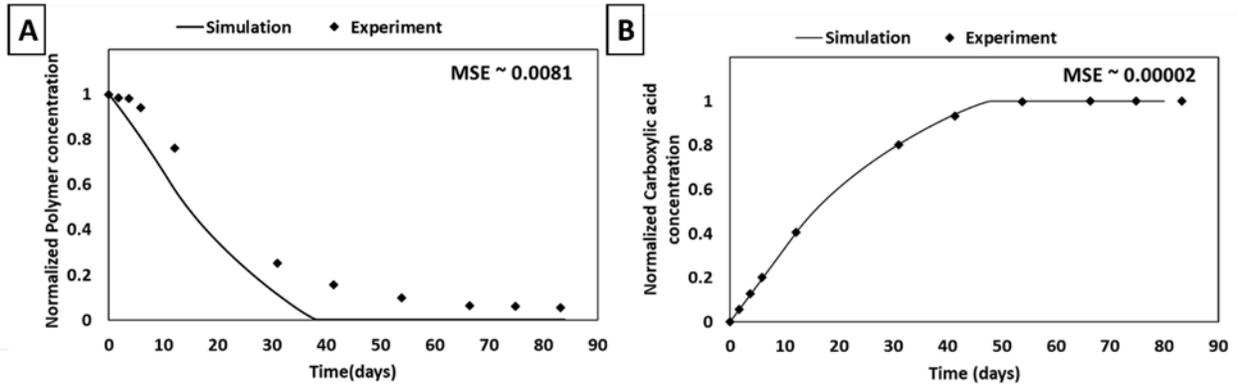

*Figure A2: Comparison between experiments (diamonds) [18] and simulations results (solid lines) during optimization procedure from kinetics model. (A) and (B) represents the normalized polymer and normalized carboxylic acid concentrations, respectively.*

The time-series scaled average concentrations of oxygen over depth is as shown in Figure A3.

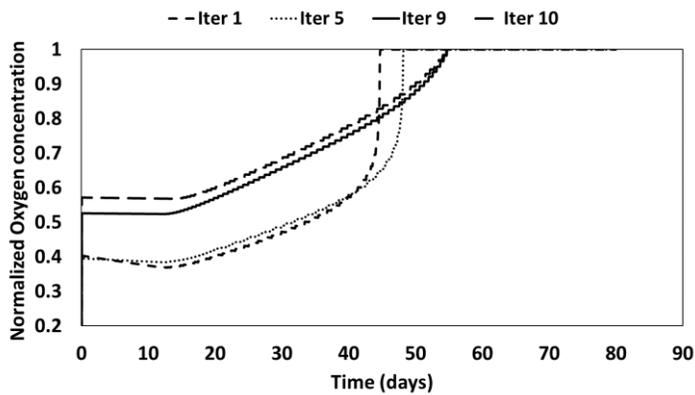

*Figure A3: Averaged oxygen concentration profiles during the optimization procedure at different iterations.*